 \documentclass[preprint,12pt]{elsarticle}
 \usepackage{amssymb}
 \usepackage{amsmath}
 \newtheorem{theorem}{Theorem}

 \newtheorem{example}{Example}

 \journal{Journal Name}

\begin{document}
	
	 \begin{frontmatter}
		
		
		%% \title{Title \tnoteref{label1}}
		%% \author{Name \corref{cor1} \fnref{label2}}
		%% \address{Address \fnref{label3}}
		
		 \title{Analytical Expression and Deconstruction of the Volume of the Controllability Ellipsoid 
		 \thanks{Work supported by the National Natural Science Foundation of China (Grant No. 61273005)}}
		
		
		 \author{Mingwang Zhao}
		
		 \address{Information Science and Engineering School, Wuhan University of Science and Technology, Wuhan, Hubei, 430081, China \\
			Tel.: +86-27-68863897 \\
		Work supported by the National Natural Science Foundation of China (Grant No. 61273005)}
		%	 \email{zhaomingwang@wust.edu.cn} 
		
		 \begin{abstract}
			In this article, we present three theorems and develop an effective analytical method to compute analytically the volume of the controllability ellipsoid for the linear discrete-time (LDT) systems with $n$ different eigenvalues. Furthermore, by deconstructing the analytical expression of the volume, some factors on the shape of the ellipsoid, the side length of its circumscribed rhomboid, the evenness of the eigenvalue distribution of the LDT system are constructed. Based on the analytical expression of the volume and these factors, the control ability can be defined, computing, analyzed, and optimized for the LDT systems.
		 \end{abstract}
		
		 \begin{keyword}
			volume computation \sep controllability ellipsoid \sep controllability Grammian matrix \sep discrete-time systems \sep state controllability
			
			
			
		 \end{keyword}
		
	 \end{frontmatter}
	
%	 \linenumbers
	
	 \section{Introduction}
	 \label{S:1}
	
	In control theory, linear discrete-time (LDT) systems can be formulated as follows:
	
	 \begin{equation}
	x_{k+1}=Ax_{k}+Bu_{k}, \quad x_{k} \in R^{n},u_{k} \in R^{r}, \label{eq:a0201}
	 \end{equation}
where $x_{k}$ and $u_{k}$ are the state variable and input
	variable, respectively, and matrices $A \in R^{n \times n}$ and $B \in
	R^{n \times r}$ are the state matrix and input matrix in the system models, respectively,
	 \cite{Kailath1980} \cite{Chen1998}. 
	To investigate the controllability of the
	LDT systems \eqref{eq:a0201}, the controllability Grammian matrix can be defined as follows
	 \begin{equation}
G_{N}= \sum _{i=0} ^{N-1} A^{i}B \left( A^{i}B \right)^T, \; N \ge n \label{eq:a0202}
 \end{equation}	
	That the rank of the Grammian matrix $G_{N}$ is $n$, the dimension of the state space of the LDT systems \eqref{eq:a0201}, is the well-known criterion on the state controllability. By the controllability Grammian matrix, the controllability ellipsoid that can describe the maximum controllable region under the total-energy constraint $ \left ( \sum _{k=0} ^{N-1} \Vert u_N \Vert ^2 _2 \le 1 \right ) $ can be defined as follows \cite{DulPag2000} \cite{KURVARA2007} \cite{PolNaKhl2008} \cite{nkh2016} \cite{cannkh2017} 
		 \begin{equation}
	E_{N}= \left \{x \left | x=G_N^{1/2}z_N, \forall z_N \in R^n: \Vert z_N \Vert _2 \le 1 \right . \right \} \label{eq:a0203}
	 \end{equation}
	
	In papers \cite{VanCari1982}, \cite{Georges1995}, \cite{PasaZamEul2014}, and \cite{Ilkturk2015}, the determinant value $ \det \left( G_{N} \right) $ and the minimum eigenvalue $ \lambda _{\textnormal {min} } \left (G_{N} \right ) $ of the controllability Grammian matrix $G_{N}$, correspondingly the volume $ \textnormal {vol} \left( E_{N} \right)$ and the minimum radius $ r _{\textnormal {min}} \left (E_{N} \right ) $ of the controllability ellipsoid $E_{N}$, can be used to quantify the control ability of the input variables to the state space, and then be chosen as the objective functions for optimizing and promoting the control ability of the dynamical systems. Due to lack of the analytical computing of the determinant $ \det \left( G_{N} \right) $ and eigenvalue $ \lambda _{\textnormal {min}} \left (G_{N} \right ) $, correspondingly the volume $\textnormal {vol} \left( E_{N} \right)$ and the radius $ r _{\textnormal {min}} \left (E_{N} \right ) $, these optimizing problems for the control ability are solved very difficulty, and few achievements about that have been made.
	
	In this paper, the analytical volume-computation of the controllability ellipsoid $E_{N}$, when the system \eqref{eq:a0201} is a sigle input system, is studied and an analytical expression on that $N \rightarrow \infty$ will be proven. By deconstructing the analytical volume expression, some factors for the ellipsoid $E_{N}$, such as, the shape factor, the minimum circumscribed rhomboid, etc, can be got. Therefore, the analytical volume and the shape factor of the ellipsoid, the minimum side length of the rhomboid can be used to describe the control ability and can be chosen as the objective functions and the constraint conditions for optimizing and promoting the control ability. Because of the analytical expression of these objective functions and constraint conditions, the optimizing problems will be solved with very effective optimizing computation.
	
 \section{The analytical volume-computing of the controllability ellipsoid for the matrix $A$ with $n$ different eigenvalues}

Based on the linear system theory \cite{Kailath1980}, \cite{Chen1998}, the LDT system \eqref{eq:a0201} can be transformed as the Jordan canonical form, and especially the LDT system \eqref{eq:a0201} with $n$ different eigenvalues can be transformed as the diagonal canonical form. 
Obviously, if the Jordan canonical form is $ \Sigma (P^{-1} A P, P^{-1}B)$ with the Jordan transformation matrix $P$, the corresponding controllability Grammian matrix can be expressed respectively as
 \begin{align} 
 \overline{G}_N
 =P^{-1}G_N P ^{-T} \label{eq:a0204}
 \end{align}
And then, the determinant $ \det \left ( \overline{G}_N \right ) $ is $ \left( \det P \right) ^{-2} \det \left ( G_N \right )$ and the volume of the controllability ellipsoid is $ \vert \det P \vert ^{-1} \textnormal {vol} \left( E_N \right) $. In this paper, the determinant and ellipsoid volume for the diagonal canonical form, respectively, are computed analytically and related results can be generalized to the general systems $ \Sigma (A,B)$.

When the system $ \Sigma (A,B)$ is a sigle-input diagonal canonical form 
as 
 \begin{align} A= \textnormal{diag} \{ \lambda_1, \lambda_2, \dots, \lambda_n \}, \; B=[b_1,b_2, \dots,b_n]^T  \label{eq:a0205}
  \end{align}
the controllability Grammian matrix can be rewritten as 
 \begin{align}
 G_N	& = \sum_{i=0}^{N-1} A^{i} B \left(A^{i}B \right)^{T} \notag \\
& = \sum_{i=0}^{N-1} \left[ \begin{array}{cccc}
b_{1}^{2} \lambda_{1}^{2i} & b_{1}b_{2} \lambda_{1}^{i} \lambda_{2}^{i} & \cdots & b_{1}b_{n} \lambda_{1}^{i} \lambda_{n}^{i} \\
b_{1}b_{2} \lambda_{1}^{i} \lambda_{2}^{i} & b_{2}^{2} \lambda_{2}^{2i} & \cdots & b_{2}b_{n} \lambda_{2}^{i} \lambda_{n}^{i} \\
 \vdots & \vdots & \ddots & \vdots \\
b_{1}b_{n} \lambda_{1}^{i} \lambda_{n}^{i} & b_{2}b_{n} \lambda_{2}^{i} \lambda_{n}^{i} & \cdots & b_{n}^{2} \lambda_{n}^{2i}
 \end{array} \right] \notag \\
& = \left[ \begin{array}{cccc}
b_{1}^{2} \frac{1- \lambda_{1}^{2N}}{1- \lambda_{1}^{2}} & b_{1}b_{2} \frac{1- \lambda_{1}^{N} \lambda_{2}^{N}}{1- \lambda_{1} \lambda_{2}} & \cdots & b_{1}b_{n} \frac{1- \lambda_{1}^{N} \lambda_{n}^{N}}{1- \lambda_{1} \lambda_{n}} \\
b_{1}b_{2} \frac{1- \lambda_{1}^{N} \lambda_{2}^{N}}{1- \lambda_{1} \lambda_{2}} & b_{2}^{2} \frac{1- \lambda_{2}^{2N}}{1- \lambda_{2}^{2}} & \cdots & b_{2}b_{n} \frac{1- \lambda_{2}^{N} \lambda_{n}^{N}}{1- \lambda_{2} \lambda_{n}} \\
 \vdots & \vdots & \ddots & \vdots \\
b_{1}b_{n} \frac{1- \lambda_{1}^{N} \lambda_{n}^{N}}{1- \lambda_{1} \lambda_{n}} & b_{2}b_{n} \frac{1- \lambda_{2}^{N} \lambda_{n}^{N}}{1- \lambda_{2} \lambda_{n}} & \cdots & b_{n}^{2} \frac{1- \lambda_{n}^{2N}}{1- \lambda_{n}^{2}}
 \end{array} \right] \label{eq:a0206}
 \end{align}
When all eigenvalues $ \lambda_i \in (-1,1), i= \overline {1,n}$, and $N \rightarrow \infty$, we have 
\begin{align}
G_ \infty 
 = \left[ \begin{array}{cccc}
 \frac{b_{1}^{2}}{1- \lambda_{1}^{2}} & \frac{ b_{1}b_{2}}{1- \lambda_{1} \lambda_{2}} & \cdots & \frac{b_{1}b_{n}}{1- \lambda_{1} \lambda_{n}} \\
 \frac{b_{1}b_{2}} {1- \lambda_{1} \lambda_{2}} & \frac{ b_{2}^{2}}{1- \lambda_{2}^{2}} & \cdots & \frac{b_{2}b_{n} }{1- \lambda_{2} \lambda_{n}} \\
\vdots & \vdots & \ddots & \vdots \\
\frac{b_{1}b_{n} }{1- \lambda_{1} \lambda_{n}} & \frac{ b_{2}b_{n} }{1- \lambda_{2} \lambda_{n}} & \cdots & \frac{ b_{n}^{2}}{1- \lambda_{n}^{2}}
\end{array} \right] \label{eq:a0207}
\end{align}

Before discussing the analytical volume computing of the infinite-time controllability ellipsoid $E_ \infty $ for the diagonal canonical form, a theorem about the determinant of the matrix $ G_ {\infty} $ is put forward and proven as follows.

 \begin{theorem} \label{th:t0201}
	For all eigenvalues $ \lambda_i \in (-1,1), i= \overline {1,n}$, we have
	 \begin{align} 
	F^{ \lambda_1, \lambda_2, \dots, \lambda_n } _ \infty&= \det \left[ \begin{array}{cccc}
	 \frac{1}{1- \lambda_{1}^{2}} & \frac{1}{1- \lambda_{1} \lambda_{2}} & \cdots & \frac{1}{1- \lambda_{1} \lambda_{n}} \\
	 \frac{1}{1- \lambda_{1} \lambda_{2}} & \frac{1}{1- \lambda_{2}^{2}} & \cdots & \frac{1}{1- \lambda_{2} \lambda_{n}} \\
	 \vdots & \vdots & \ddots & \vdots \\
	 \frac{1}{1- \lambda_{1} \lambda_{n}} & \frac{1}{1- \lambda_{2} \lambda_{n}} & \cdots & \frac{1}{1- \lambda_{n}^{2}}
	 \end{array} \right] \notag \\
	&= \left[ \prod_{1 \leq i<j \leq n} \left( \frac{ \lambda_{j}- \lambda_{i} }{ 1- \lambda_{i} \lambda_{j}} \right) ^2 \right]
	 \times \left( \prod_{i=1}^{n} \frac{1}{1- \lambda_{i}^{2}} \right) \label{eq:a0208} 
	 \end{align}
 \end{theorem}

 \textbf{Proof.} The theorem can be proven by induction method as follows.

(1) When $n=1$ and $2$, we have
 \begin{align} 
F^{ \lambda_1} _ \infty & = \det \left[
 \frac{1}{1- \lambda_{1}^{2}} \right] = \frac{1}{1- \lambda_{1}^{2}} \label{eq:a0209} \\
F^{ \lambda_1, \lambda_2} _ \infty& = \det \left[ \begin{array}{cc}
 \frac{1}{1- \lambda_{1}^{2}} & \frac{1}{1- \lambda_{1} \lambda_{2}} \\
 \frac{1}{1- \lambda_{1} \lambda_{2}} & \frac{1}{1- \lambda_{2}^{2}}
 \end{array} \right] =
 \frac{ \left( \lambda_{_2}- \lambda_{1} \right)^{2}}{ \left(1- \lambda_{1} \lambda_{2} \right)^{2} \left(1- \lambda_{1}^{2} \right) \left(1- \lambda_{2}^{2} \right) } \label{eq:a0210} 
 \end{align} 
And then, for $n=1$ and $2$, Eq. \eqref{eq:a0208} holds.

(2) It is assumed that for a given $k$, Eq. \eqref{eq:a0208} holds for $ n=k-1$, that is, we have,
	 \begin{align} 
F^{ \lambda_1, \lambda_2, \dots, \lambda_k-1 } _ \infty= \left[ \prod_{1 \leq i<j \leq k-1} \left( \frac{ \lambda_{j}- \lambda_{i} }{ 1- \lambda_{i} \lambda_{j}} \right) ^2 \right]
 \times \left( \prod_{i=1}^{k-1} \frac{1}{1- \lambda_{i}^{2}} \right) \label{eq:a0211} 
 \end{align}

(3) For $n=k$, we have
 \begin{align}
F_ \infty ^{ \lambda_1, \lambda_2, \dots, \lambda_k } & = \det \left[ \begin{array}{cccc}
 \frac{1}{1- \lambda_{1}^{2}} & \frac{1}{1- \lambda_{1} \lambda_{2}} & \cdots & \frac{1}{1- \lambda_{1} \lambda_{k}} \\
 \frac{1}{1- \lambda_{1} \lambda_{2}} & \frac{1}{1- \lambda_{2}^{2}} & \cdots & \frac{1}{1- \lambda_{2} \lambda_{k}} \\
 \vdots & \vdots & \ddots & \vdots \\
 \frac{1}{1- \lambda_{1} \lambda_{k}} & \frac{1}{1- \lambda_{2} \lambda_{k}} & \cdots & \frac{1}{1- \lambda_{k}^{2}}
 \end{array} \right] \notag \\
 & = \det \left[ \begin{array}{cccc}
 \frac{1}{1- \lambda_{1}^{2}} & 0 & \cdots & 0 \\
0 & q_{22} & \cdots & q_{2k} \\
 \vdots & \vdots & \ddots & \vdots \\
0 & q_{2k} & \cdots & q_{kk}
 \end{array} \right] \label{eq:a0212} 
 \end{align}
where
 \begin{align}
q_{22}&= \frac{1}{1- \lambda_{2}^{2}}- \frac{1}{1- \lambda_{1} \lambda_{2}} \times \frac{1- \lambda_{1}^{2}}{1- \lambda_{1} \lambda_{2}} \notag \\
&= \frac{1-2 \lambda_{1} \lambda_{2}- \lambda_{1}^{2} \lambda_{2}^{2}- \left(1- \lambda_{1}^{2}- \lambda_{2}^{2}+ \lambda_{1}^{2} \lambda_{2}^{2} \right)}{ \left(1- \lambda_{2}^{2} \right) \left(1- \lambda_{1} \lambda_{2} \right)^{2}} \notag \\
&= \frac{ \left( \lambda_{2}- \lambda_{1} \right)^{2}}{ \left(1- \lambda_{2}^{2} \right) \left(1- \lambda_{1} \lambda_{2} \right)^{2}} \label{eq:a0213} \\
q_{2k}&= \frac{1}{1- \lambda_{2} \lambda_{k}}- \frac{1}{1- \lambda_{1} \lambda_{k}} \times \frac{1- \lambda_{1}^{2}}{1- \lambda_{1} \lambda_{2}} \notag \\
&= \frac{ \left(1- \lambda_{1} \lambda_{2} \right) \left(1- \lambda_{1} \lambda_{k} \right)- \left(1- \lambda_{2} \lambda_{k} \right) \left(1- \lambda_{1}^{2} \right)}{ \left(1- \lambda_{1} \lambda_{2} \right) \left(1- \lambda_{1} \lambda_{k} \right) \left(1- \lambda_{2} \lambda_{k} \right)} \notag \\
&= \frac{1- \lambda_{1} \lambda_{2}- \lambda_{1} \lambda_{k}+ \lambda_{1}^{2} \lambda_{2} \lambda_{k}- \left(1- \lambda_{2} \lambda_{k}- \lambda_{1}^{2}+ \lambda_{1}^{2} \lambda_{2} \lambda_{k} \right)}{ \left(1- \lambda_{1} \lambda_{2} \right) \left(1- \lambda_{1} \lambda_{k} \right) \left(1- \lambda_{2} \lambda_{k} \right)} \notag \\
&= \frac{- \lambda_{1} \lambda_{2}- \lambda_{1} \lambda_{k}- \left(- \lambda_{2} \lambda_{k}- \lambda_{1}^{2} \right)}{ \left(1- \lambda_{1} \lambda_{2} \right) \left(1- \lambda_{1} \lambda_{k} \right) \left(1- \lambda_{2} \lambda_{k} \right)} \notag \\
&= \frac{ \left( \lambda_{2}- \lambda_{1} \right) \left( \lambda_{k}- \lambda_{1} \right)}{ \left(1- \lambda_{1} \lambda_{2} \right) \left(1- \lambda_{1} \lambda_{k} \right) \left(1- \lambda_{2} \lambda_{k} \right)} \label{eq:a0214} \\
& \dots \notag 
 \end{align}
And then, we have
 \begin{align}
F_ \infty ^{ \lambda_1, \lambda_2, \dots, \lambda_k } &	= \frac{1}{1- \lambda_{1}^{2}} \det \left[ \begin{array}{ccc}
 \frac{ \left( \lambda_{2}- \lambda_{1} \right)^{2}}{ \left(1- \lambda_{2}^{2} \right) \left(1- \lambda_{1} \lambda_{2} \right)^{2}} & \cdots & \frac{ \left( \lambda_{2}- \lambda_{1} \right) \left( \lambda_{k}- \lambda_{1} \right)}{ \left(1- \lambda_{1} \lambda_{2} \right) \left(1- \lambda_{1} \lambda_{k} \right) \left(1- \lambda_{2} \lambda_{k} \right)} \\
 \vdots & \ddots & \vdots \\
 \frac{ \left( \lambda_{2}- \lambda_{1} \right) \left( \lambda_{k}- \lambda_{1} \right)}{ \left(1- \lambda_{1} \lambda_{2} \right) \left(1- \lambda_{1} \lambda_{k} \right) \left(1- \lambda_{2} \lambda_{k} \right)} & \cdots & \frac{ \left( \lambda_{k}- \lambda_{1} \right)^{2}}{ \left(1- \lambda_{k}^{2} \right) \left(1- \lambda_{1} \lambda_{k} \right)^{2}}
 \end{array} \right] \notag \\
&
= \frac{1}{1- \lambda_{1}^{2}} \prod_{i=2}^{k}
 \left( \frac{ \lambda_{i}- \lambda_{1} }{ 1- \lambda_{i} \lambda_{1}} \right) ^2
 \det \left[ \begin{array}{ccc}
 \frac{1}{1- \lambda_{2}^{2}} & \cdots & \frac{1}{1- \lambda_{2} \lambda_{k}} \\
 \vdots & \ddots & \vdots \\
 \frac{1}{1- \lambda_{2} \lambda_{k}} & \cdots & \frac{1}{1- \lambda_{k}^{2}}
 \end{array} \right] \notag \\
&
= \frac{1}{1- \lambda_{1}^{2}} \prod_{i=2}^{k} \left( \frac{ \lambda_{i}- \lambda_{1} }{ 1- \lambda_{i} \lambda_{1}} \right) ^2
 \times F_ \infty ^{ \lambda_2, \lambda_3, \dots, \lambda_k } \label{eq:a0215}
 \end{align}

Therefore, by Eq. \eqref{eq:a0211} and Eq. \eqref {eq:a0215}, we have, Eq. \eqref{eq:a0208} holds for $ n=k$.

In summary, the theorem is proven by the inductive method. 
	\qed

Based on Theorem \ref{th:t0201}, the determinant of the controllability Grammian matrix and the volume of the controllability ellipsoid for the diagonal canonical form are as follows
 \begin{align} 
 \det \left ( G_ \infty \right ) &=F_N^{ \lambda_1, \lambda_2, \dots, \lambda_n } 
 \prod _{i=1}^{n} b_i^2 \label{eq:a0216} \\
\textnormal {vol} \left (E_ \infty \right) & = H_n \sqrt {\det \left ( G_ \infty \right )}
= H_n 
 \sqrt { F_N^{ \lambda_1, \lambda_2, \dots, \lambda_n } \prod _{i=1}^{n} b_i^2 } \notag \\
&= H_n \left \vert \prod_{1 \leq i<j \leq n} \frac{ \lambda_{j}- \lambda_{i} }{ 1- \lambda_{i} \lambda_{j}} \right \vert
 \times \left \vert \prod_{i=1}^{n} \frac{ b_i}{ \left(1- \lambda_{i}^{2} \right) ^{1/2} } \right \vert 
 \label{eq:a0217}
 \end{align}
where the hypersphere volume-coefficient $H_n$ and the Gamma function $ \varGamma (s)$ can be defined as 
 \begin{align} 
 H_n & = \frac{ \pi^{n/2} } { \varGamma \left( \frac {n}{2}+1 \right)}
  \label{eq:a0218}  \\
 \varGamma(s)
& = \left \{ \begin{array}{ll} 
	(s-1) \varGamma(s-1) & s>1 \\
 \sqrt \pi & s=1/2 
	 \end{array} \right.  \label{eq:a0219}
 \end{align}

According to the above computation for the diagonal canonical form, a theorem on the 
the determinant of the controllability Grammian matrix and the volume of the controllability ellipsoid for the general systems $\Sigma (A,B)$ is can be established as follows.

 \begin{theorem} \label{th:t0202}
For the LDT systems $\Sigma (A,B) $ with $n$ different eigenvalues $ \lambda_i \in (-1,1), i= \overline {1,n}$, the determinant of the controllability Grammian matrix and the volume the controllability ellipsoid for the systems can be computed analytically as follows
 \begin{align} 
 \det \left ( G_ \infty \right ) &= F_N^{ \lambda_1, \lambda_2, \dots, \lambda_n } 
\left ( \det (P) \prod _{i=1}^{n} q_iB \right) ^2 \label{eq:a0220} \\
\textnormal {vol} \left (E_\infty \right) 
&= H_n \left \vert \det (P) \prod_{1 \leq i<j \leq n} \frac{ \lambda_{j}- \lambda_{i} }{ 1- \lambda_{i} \lambda_{j}} \right \vert
\times \left \vert \prod_{i=1}^{n} \frac{ q_iB}{ \left(1- \lambda_{i}^{2} \right) ^{1/2} } \right \vert 
\label{eq:a0221}
\end{align}
where  $q_i$ is the $i$-th unit left eigenvector of the matrix $A$, and  the matrix $P$ is the diagonalization transformation matrix constructed by all unit right eigenvectors of matrix $A$.
\end{theorem}

	In papers \cite{VanCari1982}, \cite{Georges1995}, \cite{PasaZamEul2014}, and \cite{Ilkturk2015}, the determinant value $ \det \left( G_{N} \right) $  of the controllability Grammian matrix $G_{N}$ and  the volume $ \textnormal {vol} \left( E_{N} \right)$  the controllability ellipsoid $E_{N}$ are chosen as the objective functions for optimizing the control ability. 
Because lack of the analytical computing methods for $ \det \left( G_{N} \right) $  and  $ \textnormal {vol} \left( E_{N} \right)$, these optimizing problems for the control ability are solved very difficulty, and few achievements about that have been made. Based on the above analytical  computing of $\det \left ( G_ \infty \right )$ and  $\textnormal {vol} \left (E_\infty \right)$, these optimizing problems for promoting the control ability can be solved conviently and the controlled plants with the better control abilty and dynamical performance can be designed.
	
 \section{Decoding the Controllability Ellipsoid}

According to the computing equation \eqref{eq:a0221}, two factors  are deconstructed as follows.
 \begin{align} 
F_1 & = \left \vert \prod_{1 \leq i<j \leq n} \frac{ \lambda_{j}- \lambda_{i} }{ 1- \lambda_{i} \lambda_{j}} \right \vert \label{eq:a0222} \\
F_{2,i} 
&= \frac{ \left \vert q_iB \right \vert}{ \left(1- \lambda_{i}^{2} \right) ^{1/2} } 
\label{eq:a0223}
\end{align}

In fact, the above two factors can be used to describe the shape and size of the controllability ellipsoid $E_N$ and the eigenvalue evenness of the LDT system.

\subsection {The ellipsoid shape factor }

The controllability ellipsoid $E_N$ in the original space and the invariant eigen-space can be represented respectively as the following equation
 \begin{align} 
x^T \left( G_N \right) ^ {-1/2} x \le 1 
\label{eq:a0224} \\
x^T \left( \overline G_N \right) ^ {-1/2} x \le 1 
\label{eq:a0225} 
\end{align}
The $n$ radii of the ellipsoid $\overline E_N$ in the invariant eigen-space are indeed the $n$ eigenvalues 
of the Garmmian matrix $\overline G_N$, and then the shape of the ellipsoid $\overline E_N$ can be characterized by the sizes of all $n$ radii of the ellipsoid. So is the ellipsoid $G_N$.

By Eq. \eqref {eq:a0222}, we can see, when some two eigenvalues of the system matrix $A$ are approximately equal, the minimum radius of the ellipsoid $\overline E_N$ will be approximately zero, and the ellipsoid $\overline E_N$ will be flattened.
Therefore, if the distributions of all eigenvalues of the matrix $A$ are even, the ratio between the minimum and maximum radii can be avoided as a small value and then the ellipsoid $\overline E_N$ will be avoided flattened.

The factor $F_1$ deconstructed from the volume computing equation \eqref{eq:a0221} can be used to describe the evenness of the eigenvalue distribution of the Grammian matrix $G_N$ and then the uniformity of the $n$ radii of the ellipsoid $\overline E_N$. The bigger the value of the factor $F_1$, the more evenness of the eigenvalue distribution of the matrix $G_N$ in $(-1,1)$ is,  the smaller 
 the ratio between the minimum and maximum radii of the ellipsoid $\overline E_N$ is, and then the greater the volume of the ellipsoid is.
 
 	 To some extent, the minimum eigenvalue $ \lambda _{\textnormal {min} } \left (G_{N} \right ) $ of the Grammian matrix $G_{N}$  and the minimum radius $ r _{\textnormal {min}} \left (E_{N} \right ) $ of the  ellipsoid $E_{N}$ can be used to quantify the control ability of the input variables to the state space, and then be chosen as the objective functions for optimizing  the control ability of the dynamical systems in papers \cite{VanCari1982}, \cite{Georges1995}, \cite{PasaZamEul2014}, and \cite{Ilkturk2015}. Due to lack of the analytical computing methods for $ \lambda _{\textnormal {min} } \left (G_{N} \right ) $  and $ r _{\textnormal {min}} \left (E_{N} \right ) $, these optimizing problems for the control ability are solved very difficulty. In fact, the  minimum radius $ r _{\textnormal {min}} \left (E_{N} \right ) $, that is, the minimum distance between the original and the boundary of the ellipsoid, is proportional to the above shape factor $F_1$.
 The bigger the factor $F_1$ is, and  the bigger the  minimum radius $ r _{\textnormal {min}} \left (E_{N} \right ) $ is.  
 Therefore, optimizing the factor $F_1$ is equal to optimizing $ \lambda _{\textnormal {min} } \left (G_{N} \right ) $  and $ r _{\textnormal {min}} \left (E_{N} \right ) $, and then based on the above analytical expression of the fact $F_1$, optimizing control ability of the LDT systems can be carried out conveniently and effectively.

\begin{figure}[htbp]
 	\centering
 	\begin{minipage}[c]{0.45\textwidth} 
 		\centering
 		\includegraphics[width=0.8\textwidth]{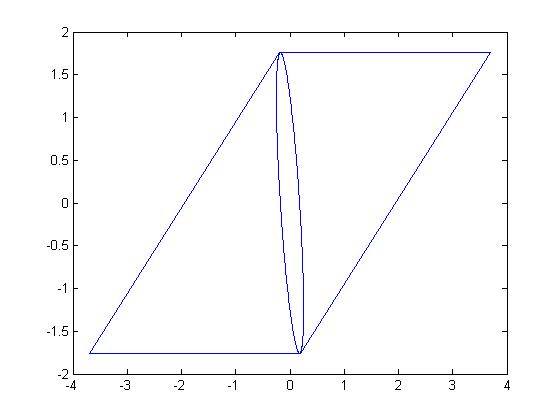} 
 		\\ (a) \footnotesize {(0.85,0.9,0.2128))}
 	\end{minipage}%	
 	\begin{minipage}[c]{0.45\textwidth}
 		\centering
 		\includegraphics[width=0.8\textwidth]{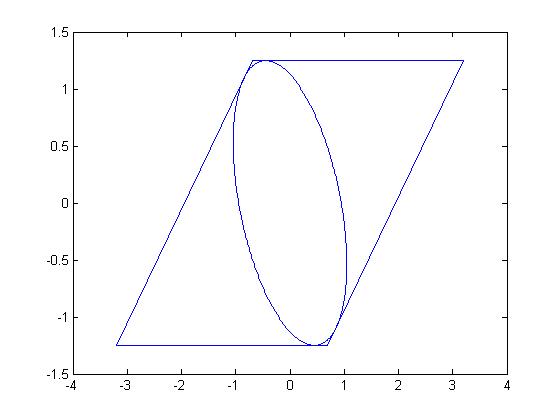}
 		\\ (b) \footnotesize {(0.6,0.9,0.6522)}
 	\end{minipage}
 	\begin{minipage}[c]{0.45\textwidth}
 		\centering
 		\includegraphics[width=0.8\textwidth]{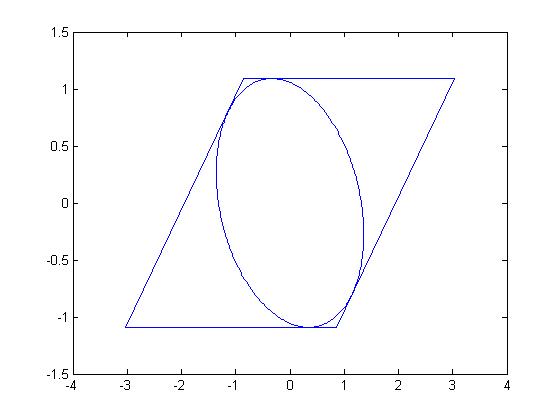}
 		%			\caption{$F_1=0.2128$ \label {fig:as03}}
 		\\ (c) \footnotesize {(0.4,0.9,0.7813)}
 	\end{minipage}
 	\caption[c]{The 2-dimensional  ellipsoid with $(\lambda_1,\lambda_2,F_1)$ \label {fig:f0201}}	
 \end{figure}

\begin{figure}[htbp]
 	\centering
 	\begin{minipage}[c]{0.45\textwidth} 
 		\centering
 		\includegraphics[width=0.8\textwidth]{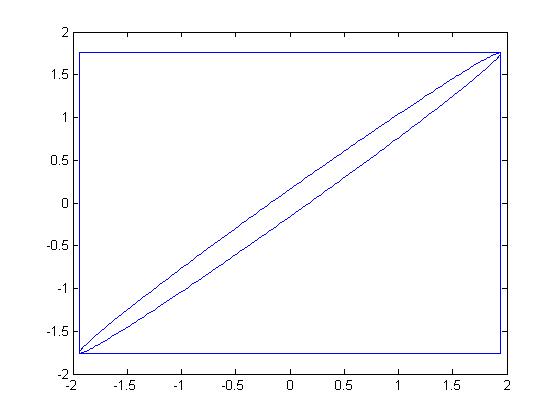}
 		\\ (a) \footnotesize {(0.85,0.9,0.2128))}
 	\end{minipage}%	
 	\begin{minipage}[c]{0.45\textwidth}
 		\centering
 		\includegraphics[width=0.8\textwidth]{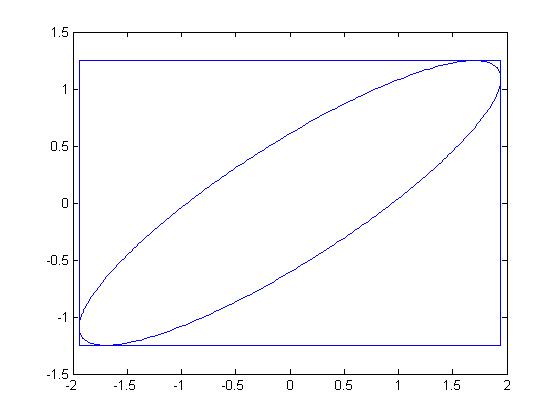}
 		\\ (b) \footnotesize {(0.6,0.9,0.6522)}
 	\end{minipage}
 	\begin{minipage}[c]{0.45\textwidth}
 		\centering
 		\includegraphics[width=0.8\textwidth]{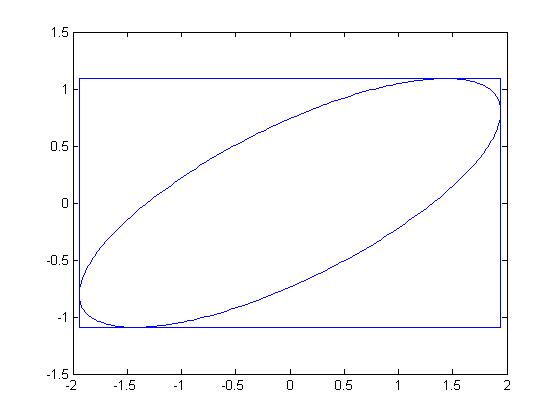}
 		%			\caption{$F_1=0.2128$ \label {fig:as03}}
 		\\ (c) \footnotesize {(0.4,0.9,0.7813)}
 	\end{minipage}
 	\caption[c]{The 2-dimensional  ellipsoid with $(\lambda_1,\lambda_2,F_1)$ in the eigen-space \label {fig:f0202}}	
 \end{figure}

 Fig. \ref{fig:f0201} shows the 2-dimensional   ellipsoids $E_{30}  $, \textit{i.e.}, the sampling number $ N=30$, generated by the 3 matrix pairs $(A,b)$ that the matrix $A$ is with the different eigenvalues and matrix $b$ is a same vector, and Fig. \ref{fig:f0202} shows the 2-dimensional   ellipsoids generated by the diagonal matrix pairs of these 3 matrix pairs $(A,b)$, \textit{that is}, the   ellipsoids in Fig. \ref{fig:f0202} are in the invariant eigen-space. These figures show us that the smaller the difference of two eigenvalues of the systems matrix $A$ is, the more flat the controllability ellipsoid is.

 %\caption[short text]{text}
 
 \subsection {The shape factors in 2-dimensional section of the ellipsoid}
 
 In fact, the shape of the ellipsoid can be observed by the shape of the each 2-dimensional section of the ellipsoid in the eigen-space. For example, for  any two eigenvalues $\lambda_i$ and $\lambda_j$, the shape factor of the 2-dimensional section $x_i-x_j$ of the ellipsoid in the eigen-space can be deconstructed from the shape factor $F_1$ as follows
  \begin{align} 
 F_{1,i,j}  = \left \vert  \frac{ \lambda_{j}- \lambda_{i} }{ 1- \lambda_{i} \lambda_{j}} \right \vert \label{eq:a0226} 
 \end{align}
 Therefore, all factor $F_{1,i,j}$ can describe the shape of the each section of the ellipsoid and then can construct the shape factor $F_1$ of  the ellipsoid.

\subsection {The eigenvalue evenness factor of the LDT system}
 
Beyond for describing the shape of the ellipsoid, the factor $F_1$ can be used to describe the evenness of the eigenvalue distribution of the LDT system $\Sigma(A,B)$. The bigger the value of the factor $F_1$ is, the more even the distribution of the $n$ eigenvalues of the matrix $A$ is, and then  the bigger the controllable region of the system is, and the stronger the control ability of the systems is.

As we know, many computing and designing methods for control laws are based on the pole assignment method with a set of given expecting closed-loop poles. How to determine the expecting closed-loop poles to obtain the greater control ability and dynamical performance for the closed-loop systems? There has been no good answer to this question. By the above analysis, we can see, optimizing the evenness factor $F_1$ of the given expecting closed-loop poles, we can get a clsode-loop system with the greater control ability and dynamical performance by the pole assignment control method.

\subsection {The circumscribed hypercube and circumscribed rhomboid}

 The factor $F_{2,i}$ is indeed the biggest distance of the boundary of the ellipsoid $\overline E_N$ in the $i$-dimensional space. In fact, the $n$ side lengths of the circumscribed hypercube of the ellipsoid $\overline E_N$, shown in Fig. \ref{fig:f0202},  are $2F_{2,i},i=\overline{1,n}$, and then the volume of circumscribed hypercube is the production  $2^n \prod _{i=1} ^{n} F_{2,i}$  among the all factor $ F_{2,i}$.  By the volume equation \eqref{eq:a0221}, the volume of the ellipsoid can be represented as the volume of the circumscribed hypercube, the shape factor $F_1$, and some constant coefficient.
 
 Therefore, we have the following discussions:
 
 (1) When the circumscribed hypercubes of two controllability ellipsoids are approximated, the bigger the shape factor $F_1$ is , the bigger the volume of the ellipsoid is. Furthermore, the almost all  2-dimensional shape factors $F_{1,i,j}$ for some ellipsoid are bigger than the another, the whole ellipsoid can be said to be bigger than the another.
 
  (2) When the shape factors $F_1$ of two controllability ellipsoids are approximated, the bigger  the circumscribed hypercubes  is, the bigger the volume of the ellipsoid is. Furthermore, the almost all  2-dimensional shape factors $F_{1,i,j}$ for the two ellipsoids are approximated, the whole  ellipsoid with the  bigger circumscribed hypercubes  
  can be said to be bigger than the another.
   
 \section{Analytic Volume-Computation for the systems with the complex eigenvalues}

 \textbf{Theorem \ref{th:t0202}} for the LDT systems with the real eigenvalues can be generalized to the LDT systems with the complex eigenvalues, and the corresponding the theorem can be stated as follows.
 
 \begin{theorem} \label{th:t0203}
 	When all $n$ different complex eigenvalues $ \lambda_i \left( i= \overline {1,n} \right) $ of the LDT systems $\Sigma (A,B) $ satisfy that $ \vert \lambda_i \vert \in [0,1)$, the determinant of the controllability Grammian matrix and the volume the controllability ellipsoid for the systems can be computed analytically as follows
 	\begin{align} 
 	\det \left ( G_ \infty \right ) &=  \left[ \det (P)  \prod_{1 \leq i<j \leq n}  \frac{ \lambda_{j}- \lambda_{i} }{ 1- \lambda_{i} \lambda_{j}}   \right] ^2
 \left[\prod_{i=1}^{n} \frac{ \left (  q_iB \right) ^2 }{1- \left \vert \lambda_{i} \right \vert ^{2}} \right] \label {eq:a0227}
 \\ 	
 	\textnormal {vol} \left (E_\infty \right) 
 	&= H_n \left \vert \det (P) \prod_{1 \leq i<j \leq n} \frac{ \lambda_{j}- \lambda_{i} }{ 1- \lambda_{i} \lambda_{j}} \right \vert
 	\times \left \vert \prod_{i=1}^{n} \frac{ q_iB}{ \left(1- \left \vert \lambda_{i} \right \vert ^{2} \right) ^{1/2} } \right \vert 
 	\label{eq:a0228}
 	\end{align}
 	where the complex vector $q_i$ is the $i$-th unit left eigenvector of the matrix $A$, and the complex matrix $P$ is the diagonalization transformation matrix constructed by all unit right eigenvector of the matrix $A$.
 \end{theorem}

Similar to Section 3, based on these analytical expression, the shape factors of the controllability ellipsoid and the evenness factor of the eigenvalue distribution for LDT systems with complex eigenvalues can be got.

According to the computing equation \eqref{eq:a0228}, some factors described the shape and size of the controllability ellipsoid are deconstructed as follows.
\begin{align} 
F_1 & = \left \vert \prod_{1 \leq i<j \leq n} \frac{ \lambda_{j}- \lambda_{i} }{ 1- \lambda_{i} \lambda_{j}} \right \vert \label{eq:a0229} \\
F_{2,i} 
&= \frac{ \left \vert q_iB \right \vert}{ \left(1- \left \vert \lambda_{i} \right \vert ^{2} \right) ^{1/2} } 
\label{eq:a0230}
\end{align}
Similar to the factors for the matrix $A$ with the real eigenvalues in Eqs. \eqref{eq:a0222} and \eqref{eq:a0223},  the above factors can be describe the shape and size of the controllability ellipsoid $E_N$ and the eigenvalue evenness factor of the LDT systems.
	
For the conjugate complex eigenvalue pair $\left( \lambda_i,\lambda_{i+1} \right)=\left( \lambda_i,\lambda_i^* \right)$, the shape factor of the  2-dimensional section of the ellipsoid in the eigen-space is
  \begin{align} 
  F_{1,i,i+1}  = \left \vert \frac{ \lambda_{i}^*- \lambda_{i} }{ 1- \lambda_{i} \lambda_{i}^*} \right \vert =  \frac{ 2\textnormal{Im}\lambda_{i} }{ 1- \left \vert \lambda_{i} \right \vert ^2} \label{eq:a0231}
  \end{align}

\begin{figure}[htbp]
	\centering
	\begin{minipage}[c]{0.6\textwidth} 
		\centering
		\includegraphics[width=0.8\textwidth]{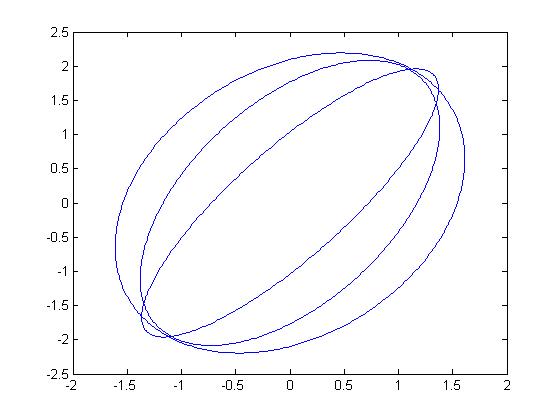} %0.5
	\end{minipage}%	
	\caption[c]{The 2-dimensional  ellipsoid with  the complex eigenvalues }	
	\label {fig:as05}
\end{figure}

Fig. \ref{fig:as05} shows the 2-dimensional ellipsoids $E_N (30) $ generated by the following matrix pairs $(A,b)$ with the complex eigenvalues
\begin{align}
A=\left[ \begin{array}{cc}
0.8 & -a \\
a & 0.8
\end{array} \right],\;\;
b=\left[  \begin{array}{c} 1 \\ 1 
\end{array} \right] \label{eq:a0232}
\end{align}
where $a=0.1,0.2,0.3$,  the corresponding the eirenvalues are $0.8\pm0.1i,0.8\pm0.2i,0.8\pm0.3i$, and the factors $F_1=1.032,2.282,4.121$. By Fig.  \ref{fig:as05}, we can see, the bigger the image part of the conjugate complex eigenvalue pair is, the greater the volume of the ellipsoid.
  
 \section {Numerical Experiments}
 
 \begin{example} \label{ex:e0201}
 	Computing the volume, shape factor, and the side lengths of the circumscribe hypercube of the controllability ellipsoid generated by the following matrix pair
 	\begin{align}
 	(A,b)=\left( \left[ \begin{array}{ccc}
 	0 & 1 & 0 \\
 	0 & 0 & 1 \\
 	0.432 & -1.74 & 2.3 \end{array} \right],
 	\left[ \begin{array}{c}
 	0 \\
 	0 \\
 	1 \end{array} \right]
 	\right) \label{eq:a0233}
 	\end{align} 
 	\end{example}

By the diagonal matrix transformation, the diagonal matrix pair is as follows
 	\begin{align}
\left (\hat A, \hat b \right )=\left( \textnormal {diag} \{ 0.6,0.8,0.9 \}, [    2.034,7.158,5.235]^T \right) \label{eq:a0234}
\end{align} 
and then, the analytical computing results about the volume and these factors are as \textbf{Table \ref{tab:ab02}}

	\begin{table}
	\centering
	\caption{Computing results of the volume and shape factors for the systems $\Sigma(A,B)$ }
	\label{tab:ab02} % Give a unique label
	\begin{tabular}{rl}
		\hline \noalign{ \smallskip}
		factors & values \\
		\hline
		\noalign{ \smallskip}
volume: & 298.8566 \\
shape factor $F_1$: & 0.0896 \\
2-D shape factors $F_{1,1,2},F_{1,1,3},F_{1,2,3}$: & 0.3846, \; 0.6522, \; 0.3571 \\
the side length $F_{2,1},F_{2,2},F_{2,3}$: & 2.5425,\; 11.9300,\; 12.0099\\
		\noalign{ \smallskip} \hline
	\end{tabular}
\end{table}

\begin{figure}[htbp]
	\centering
	\begin{minipage}[c]{0.95\textwidth} 
		\centering
		\includegraphics[width=0.9\textwidth]{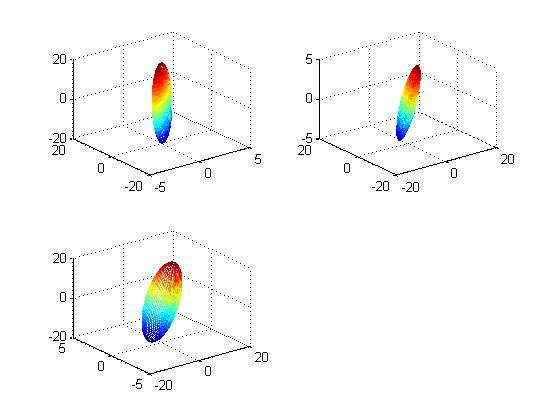} %0.5
	\end{minipage}%	
	\caption[c]{The 3-dimensional  ellipsoid in the eifen-sapce }	
	\label {fig:f0204}
\end{figure}

Fig. \ref{fig:f0204} shows the 3-dimensional ellipsoids $E_N (60) $ generated by the above matrix pairs $\left (\hat A, \hat b \right )$  in the eigen-space, from three visual angle, and these 3 figures show us the 3 sections of the ellipsoid.

 \section {Conclusions}
 
 			In this article, we present three theorems
 and develop an effective analytical method to compute the volume of the controllability ellipsoid for the LDT systems with $n$ different eigenvalues.
 Furthermore, by deconstructing the analytical expression of the volume, some factors on the shape of the ellipsoid, the side length of its circumscribed rhomboid, the evenness of the eigenvalue distribution of the LDT system are constructed. Based on the analytical expression of the volume and these factors, the control ability can be defined, computing, analyzed, and optimized.

	 \bibliographystyle{model1b-num-names}
 \bibliography{zzz}

\begin{thebibliography}{11}
\expandafter\ifx\csname natexlab\endcsname\relax\def\natexlab#1{#1}\fi
\providecommand{\bibinfo}[2]{#2}
\ifx\xfnm\relax \def\xfnm[#1]{\unskip,\space#1}\fi
%Type = Article
\bibitem[{Canfield and Nkhumise(2017)}]{cannkh2017}
\bibinfo{author}{S.L. Canfield}, \bibinfo{author}{R.~Nkhumise},
  \bibinfo{title}{Controllability ellipse to evaluate performance of mobile
  manipulators for manufacturing tasks}, \bibinfo{journal}{Journal of
  Mechanisms and Robotics} \bibinfo{volume}{9} (\bibinfo{year}{2017}).
%Type = Book
\bibitem[{Chen(1998)}]{Chen1998}
\bibinfo{author}{C.T. Chen}, \bibinfo{title}{Linear system theory and design},
  \bibinfo{publisher}{Oxford University Press, Inc. New York, NY, USA},
  \bibinfo{edition}{3rd} edition, \bibinfo{year}{1998}.
%Type = Book
\bibitem[{Dullerud and Paganini(2000)}]{DulPag2000}
\bibinfo{author}{G.E. Dullerud}, \bibinfo{author}{F.~Paganini},
  \bibinfo{title}{A Course in Robust Control Theory: A Convex Approach},
  \bibinfo{publisher}{Sringer}, \bibinfo{year}{2000}.
%Type = Inproceedings
\bibitem[{Georges(1995)}]{Georges1995}
\bibinfo{author}{D.~Georges}, \bibinfo{title}{The use of observability and
  controllability gramians or functions for optimal sensor and actuator
  location in finite-dimensional systems}, in: \bibinfo{booktitle}{Proc. of
  IEEE Conf. on Decision and Control, New Orleans, LA, USA}, p.
  \bibinfo{pages}{3319–3324}.
%Type = Phdthesis
\bibitem[{Ilkturk(2015)}]{Ilkturk2015}
\bibinfo{author}{U.~Ilkturk}, \bibinfo{title}{Observability Methods in Sensor
  Scheduling}, Ph.D. thesis, ARIZONA STATE UNIVERSITY, \bibinfo{year}{2015}.
%Type = Book
\bibitem[{Kailath(1980)}]{Kailath1980}
\bibinfo{author}{T.~Kailath}, \bibinfo{title}{Linear systems},
  \bibinfo{publisher}{Prentice-Hall}, \bibinfo{address}{Englewood Cliffs, NJ},
  \bibinfo{year}{1980}.
%Type = Article
\bibitem[{Kurzhanskiy and Varaiya(2007)}]{KURVARA2007}
\bibinfo{author}{A.~Kurzhanskiy}, \bibinfo{author}{P.~Varaiya},
  \bibinfo{title}{Ellipsoidal techniques for reachability analysis of
  discrete-time linear systems}, \bibinfo{journal}{IEEE Trans. on Automatic
  Control} \bibinfo{volume}{52} (\bibinfo{year}{2007}) \bibinfo{pages}{26--38}.
%Type = Masterthesis
\bibitem[{Nkhumise(2016)}]{nkh2016}
\bibinfo{author}{R.M. Nkhumise}, \bibinfo{title}{Controllability Ellipse -- a
  Method to Eavluate Performance of Mobile Manipulators Applied to Welding},
  Master's thesis, Tennessee Technological University, \bibinfo{year}{2016}.
%Type = Article
\bibitem[{Pasqualetti et~al.(2014)Pasqualetti, Zampieri and
  Bullo}]{PasaZamEul2014}
\bibinfo{author}{F.~Pasqualetti}, \bibinfo{author}{S.~Zampieri},
  \bibinfo{author}{F.~Bullo}, \bibinfo{title}{Controllability metrics,
  limitations and algorithms for complex networks}, \bibinfo{journal}{IEEE
  Trans. on Control of Network Systems} \bibinfo{volume}{1}
  (\bibinfo{year}{2014}) \bibinfo{pages}{40--52}.
%Type = Inproceedings
\bibitem[{Polyak et~al.(2008)Polyak, Nazin and Khlebnikov}]{PolNaKhl2008}
\bibinfo{author}{B.~Polyak}, \bibinfo{author}{S.~Nazin},
  \bibinfo{author}{M.~Khlebnikov}, \bibinfo{title}{The invariant ellipsoid
  technique for analysis and design of linear control systems}, in:
  \bibinfo{booktitle}{Advances in Mechanics: Dynamics and Control: Proceedings
  of the 14th International Workshop on Dynamics and Control, ed. F.L.
  Chernousko, G.V. Kostin, V.V. Saurin, Moscow: Nauka,}, pp.
  \bibinfo{pages}{239--246}.
%Type = Techreport
\bibitem[{VanderVelde and Carignan(1982)}]{VanCari1982}
\bibinfo{author}{W.~VanderVelde}, \bibinfo{author}{C.~Carignan},
  \bibinfo{title}{A dynamic measure of controllability and observability for
  the placement of actuators and sensors on large space structures},
  \bibinfo{type}{Technical Report}, NASA-CR-168520, SSL-2-82,
  \bibinfo{year}{1982}.

\end{thebibliography}
 \end{document}